\documentclass[manuscript]{emulateapj-rtx4}

\def\icarus{Icarus}

\slugcomment{To appear in ApJ}

\shorttitle{The survival of gas giant planets}

\begin{document}

\title{The migration of gas giant planets in gravitationally unstable discs}

\author{Dimitris Stamatellos}
\affil{Jeremiah Horrocks Institute for Mathematics, Physics \& Astronomy\\University of Central Lancashire, Preston, PR1 2HE, UK}

\email{dstamatellos@uclan.ac.uk}

\begin{abstract}
Planets form in the discs of gas and dust that surround young stars. It is not known whether gas giant planets on wide orbits form the same way as Jupiter or by fragmentation of gravitationally unstable discs. Here we show that a giant planet, which has formed in the outer regions of a protostellar disc, initially migrates fast towards the central star (migration timescale $\sim 10^4$ yr) while accreting gas from the disc. However, in contrast with previous studies, we find that the planet eventually opens up a gap in the disc and the migration is essentially halted. At the same time, accretion-powered radiative feedback from the planet, significantly limits its mass growth, keeping it  within the planetary mass regime (i.e. below the deuterium burning limit) at least for the initial stages of disc evolution. Giant planets may therefore be able to survive on wide orbits despite their initial fast inward migration, shaping the environment in which terrestrial planets that may harbour life form.
\end{abstract}

\keywords{protoplanetary disks -- planet-disk interactions --  planets and satellites: formation, gaseous planets -- hydrodynamics}

\section{Introduction}
Observations of exoplanetary systems are important as they provide significant clues about the formation of our Solar system.  In the last few years direct imaging of  massive giant planets on wide orbits has become possible \citep{Marois:2008a, Ireland:2011a,Kuzuhara:2013a, Aller:2013a,Bailey:2014a,Rameau:2013a,Galicher:2014a, Kraus:2014a}. These planets challenge our understanding of planet formation as they are unlikely to form by core accretion, i.e. the coagulation of dust particles to progressively larger aggregates and an accretion of a gaseous envelope  \citep{Safronov:1969a,Goldreich:1973a,Mizuno:1980a,Pollack:1996a}. Gravitational fragmentation of protostellar discs is an alternative formation scenario \citep{Kuiper:1951a,Cameron:1978a, Boss:1997a, Boley:2009a}. It has been argued that protoplanets forming  in gravitationally unstable  discs  quickly migrate towards the host star \citep{Baruteau:2011a, Michael:2011a}, where they may be disrupted by tidal torques \citep{Nayakshin:2013a, Tsukamoto:2013c}. Moreover, it has been suggested that protoplanets rapidly grow in mass by accreting material from the disc to become brown dwarfs and low-mass stars \citep{Rafikov:2005a,Stamatellos:2009a,Zhu:2012a}. 

In this Letter we present models of the interaction  of a newly formed massive planetary embryo (henceforth {\it protoplanet}) with a protostellar disc that contradict the above claims. We critically include in our models two ingredients that their combined effect has been ignored by previous studies: (i) gas accretion onto the protoplanet, and (ii) accretion-powered radiative feedback  from the protoplanet.

\section{Computational model}
We assume that the protoplanet has formed in the protostellar disc of a Sun-like star by gravitational fragmentation.  For this to happen the disc has to be relatively massive and the protoplanet's initial orbital radius is at least $50-100$ AU, as the conditions are favourable for fragmentation only in the disc outer region \citep{Rafikov:2005a,Stamatellos:2009a}. The mininum initial mass of the protoplanet is set by the opacity limit for fragmentation which is thought to be $1-5$~M$_{\rm J}$, where M$_{\rm J}$ is the mass of Jupiter \citep{Low:1976a, Whitworth:2006a,Boley:2010b,Kratter:2010b}.  Here we examine the evolution of such a protoplanet as it interacts with its parent disc, using a 3D  Smoothed Particle Hydrodynamics method that includes the effects of radiative transport in the disc \citep{Stamatellos:2007b,Stamatellos:2009d}. 

\subsection{Initial conditions}
We assume a star-disc system in which the central  star  has initial mass $M_\star=1\,{\rm M}_{\sun}$.  The initial disc mass is  $M_{_{\rm D}}=0.1~{\rm M}_{\sun}$ and the initial disc radius is $R_{_{\rm D}}=100$~AU. The disc is represented by $10^6$ SPH particles.  The inner disc boundary is set at  0.2~AU. The disc initial surface density is 
\begin{equation}
\Sigma_{_0}(R)=\Sigma(1 {\rm AU})\,\left(\frac{R}{\rm AU}\right)^{-1}\,,
\end{equation}
and the disc temperature
\begin{equation}\label{EQN:TBG}
T_{_0}(R)=250\,{\rm K}\,\left(\frac{R}{\rm AU}\right)^{-3/4}+10~{\rm  K}\,,
\end{equation}
where $\Sigma(1 {\rm AU})$ is determined by the disc mass and radius, and $R$ is the distance from the central star measured on the disc midplane. The initial temperature profile corresponds to a flat disc that reprocesses stellar radiation and it is steeper when compared with observations (e.g. \cite{Andrews:2009a} find a temperature profile  $R^{-q}$; $q\approx 0.4-0.74$, for discs in Ophiuchus). However the actual temperature profile attained once the disc is allowed to evolve  is less steep due to additional heating sources (viscous and accretion heating).

We let the disc to relax  for 3~kyr  ($\sim3$ outer orbital periods) and we then embed in it a protoplanet of initial mass $M_{p,i}=1~{\rm M_{\rm J}}$ at distance  $R_{p,i}=$ 50~AU from the central star and let the disc-planet system to evolve. The protoplanet's initial  velocity is set the same as the velocity of the local gas, i.e. Keplerian (including the contribution from the disc mass within the protoplanet's orbit). We  assume an initial circular orbit (eccentricity $e_i=0$). The protoplanet is allowed to accrete gas  from the disc, if this gas approaches within $0.1$~AU from the protoplanet and is bound to it. This distance is always much smaller  than the Hill radius of the protoplanet, defined as the region where the protoplanet's gravity dominates over the gravity of the central star.  Therefore, the region around the protoplanet is resolved appropriately.
 
 \subsection{Hydrodynamics \& radiative transfer}

We use the SPH code {\sc seren} \citep{Hubber:2011a} to treat the gas thermodynamics. The code  invokes an octal tree (to compute gravity and find neighbours), multiple particle timesteps, and a 2$^{\rm nd}$-order Runge-Kutta integration scheme. The disc self-gravity is therefore included in the simulations. The code uses time-dependent artificial viscosity with parameters $\alpha_{\min}=0.1$, $\alpha_{\max}=1$ and $\beta=2\alpha$,  so as to reduce artificial shear viscosity. The chemical and radiative processes that regulate the gas temperature are treated with the approximation of \cite{Stamatellos:2007b} \citep[see also][]{Forgan:2009b}. We adopt opacity tables appropriate for protostellar discs \citep{Semenov:2003a}.

\subsection{Star and protoplanet representation.}

The central star and the protoplanet are represented by sink particles that interact with the rest of the computational domain only through their gravity (and luminosity, when irradiation from the protoplanet is taken into account). The sink radius of the central star is set to $R_{\rm sink,\star}=0.2$~AU, and the sink radius of the protoplanet is set to $R_{\rm sink,p}=0.1$~AU. This value is always much smaller (by a at least a factor of $\sim 25$) than the Hill radius of the protoplanet, defined as  the region where the gravity of the protoplanet dominates over the gravity of the star, i.e.
$ R_{\rm sink, p}<R_{\rm H}=R\left({M_p}/{3M_\star}\right)^{1/3}$. 
The Hill radius increases as the protoplanet accretes material from the disc or it decreases as the protoplanet moves closer to the central star.  Gas particles accrete onto a star or protoplanet sink when they are within the sink radius and bound to the sink.

\subsection{Radiative feedback from the star and the protoplanet.}

The radiation feedback from the star and the protoplanet  is taken into account by invoking a pseudo-ambient radiation field with temperature $T_{_{\rm A}}({\bf r})$  that is a function of the position relative to the star and the protoplanet \citep{Stamatellos:2007b,Stamatellos:2011a,Stamatellos:2012a}.  This temperature effectively sets the minimum temperature that the gas can attain when cooled radiatively.
The contribution to  $T_{_{\rm A}}({\bf r})$ from the central star is set to 
 \begin{eqnarray}
T_{_{\rm A}}^{\star}({\bf r})&=&250\,{\rm K}\,\left(\frac{R}{\rm AU}\right)^{-3/4}+10~{\rm  K}\,,
\end{eqnarray}
where $R$ is the distance from the star measured on the disc midplane. The contribution to  $T_{_{\rm A}}({\bf r})$ from the protoplanet is 
 \begin{eqnarray}
T_{_{\rm A}}^{\rm planet}({\bf r})&=&\left(\frac{L_p}{16\,\pi\,\sigma_{_{\rm SB}}\,|{\bf r}-{\bf r}_p|^2}\right)^{1/4}\,,
\end{eqnarray}
where $L_p$ and  ${\bf r}_p$, are the luminosity and position of the  protoplanet, respectively. This luminosity  is given by
\begin{equation}
L_p=f\frac{G M_p \dot{M}_p}{R_{\rm acc}}\,,
\end{equation}
where $M_p$ is the mass of the protoplanet,  $\dot{M}_p$ is the accretion rate on to it, and $R_{\rm acc}$ the accretion radius.  $f=0.75$ is the fraction of the accretion energy that is radiated away at the surface  of the protoplanet, rather than being expended driving jets and/or winds \citep{Machida:2006a}. As we assume that this planet has formed by gravitational instabilities in the disc, the accretion happens onto the second hydrostatic core. The radius of the second core
is uncertain; it is estimated to be $\sim 1-20~{\rm R}_{\sun}$ \citep{Masunaga:2000a, Tomida:2013a, Vaytet:2013a}. Here we shall assume $R_{\rm acc}=1{\rm R}_{\sun}$.
 The total pseudo-ambient temperature is 
 \begin{eqnarray}
T_{_{\rm A}}^4({\bf r})&=&\left[T_{_{\rm A}}^{\star}({\bf r})\right]^4+\left[T_{_{\rm A}}
^{\rm planet}({\bf r})\right]^4\,.
\end{eqnarray}
We note that the radiative feedback from the star is fixed whereas the radiative feedback from the protoplanet is variable with time and depends on the accretion of gas onto it.

\section{Migration of gas giant planets in gravitationally unstable discs}

\begin{figure}[t]
\centerline{
\includegraphics[width=7cm]{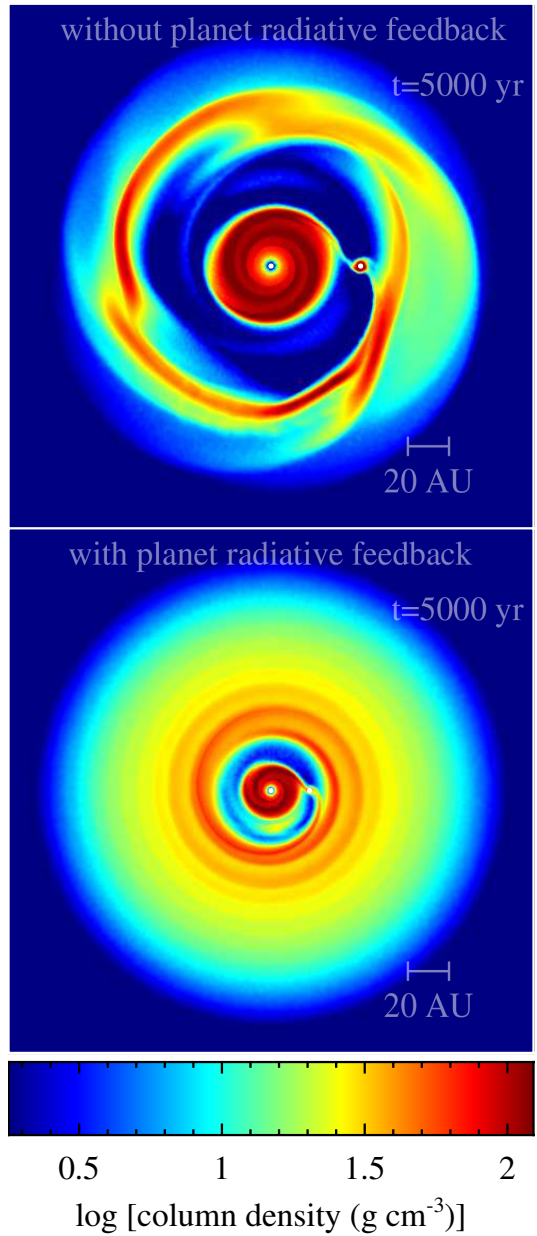}}
\caption{Column density plots showing gap opening induced by a Jupiter-mass protoplanet in  an 0.1-M$_{\odot}$ disc, in two simulations: without (top) and with (bottom) radiative feedback from the protoplanet.  The star (in the centre) and the protoplanet are depicted by thick white dots. The protoplanet in the first simulation opens up a deep, wide gap and grows to become a brown dwarf, migrating initially inwards and subsequently outwards. In the simulation with radiative feedback the gap is shallow and narrow. The protoplanet migrates inwards but its mass growth is suppressed, so that it becomes a wide-orbit planet.}
\label{fig:natsnapshots}
\end{figure}

\begin{figure}[t]
\centerline{\includegraphics[width=10cm]{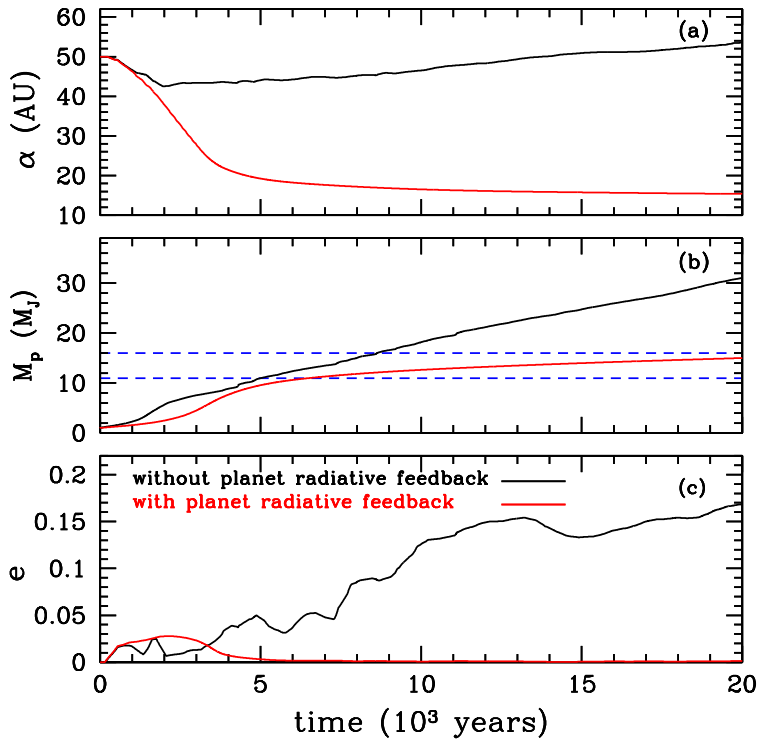}}
\caption{The evolution of {\bf (a)} the semi-major axis $\alpha$, {\bf (b)} the mass $M_{\rm p}$, and {\bf (c)} the eccentricity $e$ of the protoplanet. 
The protoplanet initially migrates fast inwards, but it grows in mass and it is able to open up a gap in the disc. The migration then is essentially halted. In the non-radiative case the protoplanet  becomes a brown dwarf, whereas in the case with radiative feedback  mass growth is suppressed and  the protoplanet's mass remains within the planetary-mass regime (marked by the horizontal dashed lines in (b)). Radiative feedback keeps the protoplanet on an circular orbit ($e\approx0$).}
\label{fig:natevolution}
\end{figure}

\begin{figure}[t]
\centerline{\includegraphics[width=9cm]{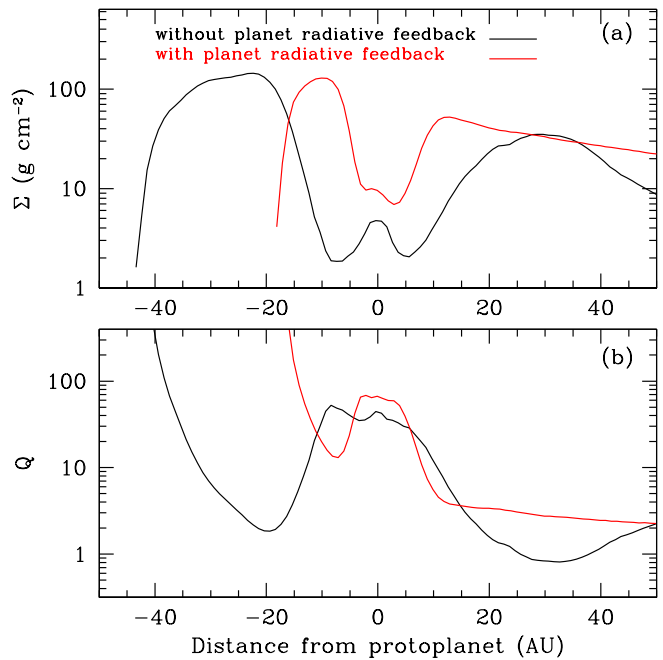}}
\caption{(a) Azimuthally averaged surface density of the disc in the protoplanet neighbourhood (at $5$~kyr ), for the two cases examined here (without/with irradiation from the protoplanet).  The radial cells where the surface density is averaged are centred around the star, but the distance in the graph is given with respect to the cell of the protoplanet (to make comparison easier). The gap is shallower and narrower when radiative feedback from the protoplanet is taken into account. (b) As above, but for the azimuthally averaged Toomre parameter $Q(R)={c(R)\,\Omega(R)}/{\pi\,G\,\Sigma(R)}$, where $c$ is  the midplane isothermal sound speed, $\Omega$ is the the angular velocity, $\Sigma$ is the surface density, and $R$ is the distance from the star. Radiative feedback from the protoplanet stabilises the disc ($Q>1$).}
\label{fig:natgap}
\end{figure}

We investigate the planet-disc interactions and the evolution of the properties of the protoplanet in two cases, with and without radiative feedback from the protoplanet (Figure~\ref{fig:natsnapshots}).  The initial evolution of the protoplanet orbital parameters and mass shows similar pattern in both cases (Figure~\ref{fig:natevolution}). The protoplanet initially migrates inwards fast, with a migration timescale of $\sim10^4$~years, in agreement with previous studies \citep{Baruteau:2011a,Michael:2011a}. As the protoplanet migrates inwards,  its mass increases significantly by accreting gas from the disc. Eventually the protoplanet is massive enough to be able to open up a gap and the migration either slows down (when the radiative feedback from the protoplanet is taken into account; migration timescale $\sim10^5$~years) or even changes to a slight outward migration (for the case without radiative feedback from the protoplanet; outward migration timescale $\sim10^5$~years). After the gap  opens up, the protoplanet continues to grow in mass but rather slowly, mainly by accreting material from its circumplanetary disc, a disc formed within the Hill sphere of the protoplanet. The time required for opening up a gap is different between the two cases that we present here. In the radiative feedback case the radiation emitted from the protoplanet heats  the disc, the opening up of  the gap is more difficult and takes twice as much time to happen ($4\times10^4$~yr in comparison with $2\times10^4$~yr for the non-radiative  protoplanet). The gap not only takes more time to open but it is  also shallower and less wide (Figure~\ref{fig:natgap}a), as the protoplanet moves closer to the central star (where both the disc scale height and Hill radius of the protoplanet are smaller).  The protoplanet grows in mass slower, and the inward migration  continues. The protoplanet in this case migrates closer to the central star (Figure~\ref{fig:natevolution}a) but its mass growth is considerably suppressed (Figure~\ref{fig:natevolution}b). At the same time the orbit of the protoplanet is circularized (Figure~\ref{fig:natevolution}c).

The final outcome of disc-planet interactions is distinctly different for the two cases presented here (Figure~\ref{fig:natsnapshots}). In the case without radiative feedback the protoplanet eventually becomes a brown dwarf with a mass of 28~M$_{\rm J}$, on a  eccentric orbit ($e\approx0.17$), and a semi-major axis of 50~AU. When irradiation from the accreting protoplanet is included in the model, the protoplanet reaches a mass of only 14~M$_{\rm J}$, i .e. around the deuterium burning limit and  within the planetary-mass regime \citep[$<11-16.3$~M$_{\rm J}$;][]{Spiegel:2011a}, while it remains on a relatively wide (semi-major axis of 15~AU), circular orbit, avoiding excessive inward migration.

The role of the radiative feedback from the protoplanet is critical. Initially, irradiation from the protoplanet heats the disc making the opening up of the gap more difficult, while thermal pressure delays  large gas accretion onto the protoplanet. Therefore,  the protoplanet continues its inward migration without accreting excessively. Its migration is eventually slowed down once the gap is opened up but the protoplanet still continues to heat and stabilize the disc. The effective viscosity due to the disc self-gravity is small and the inward flow of gas is relatively slow. Most of the gas within the protoplanet orbital radius is accreted onto the central star creating a inner hole in the disc. On the other hand when radiative feedback from the protoplanet is not  taken into account, the disc remains unstable; the Toomre parameter Q at the edges of the gap is below 1 (see Figure~\ref{fig:natgap}b). Therefore, the gas flows rather fast towards the outer edge of the gap, onto the circumplanetary disc and eventually onto the protoplanet, as the effective viscosity due to the disc self-gravity is large. The gravitationally unstable gap edges may  also be responsible for the outward migration of the protoplanet \citep{Lin:2012b}.

These results contradict previous findings in which inward migration is fast and continues towards the central star as the protoplanet is not able to open up a gap \citep{Baruteau:2011a,Michael:2011a}, questioning whether the survival of giant planets formed early on during the lifetime of a disc by fragmentation is possible. The critical difference of this work with previous models is that the protoplanet is allowed to grow in mass and therefore becomes massive enough to be able to open up a gap and its migration is slowed down.


\section{Numerical resolution and comparison with previous studies.}

To test our numerical code we performed simulations of disc-planet interactions in a system where a Jupiter-mass planet is embedded in a low-mass (0.005~M$_{\odot}$) disc at a circular orbit at a distance of  5.2~AU from an 1-M$_{\odot}$ star. This  problem has been studied both numerically, with grid-based \citep{Bate:2003a} and particle-based \citep{Ayliffe:2009a} codes, and analytically \citep{Ward:1997a}. We performed simulations matching the initial conditions of previous authors \citep{Ayliffe:2009a} using $10^6$ and $2\times10^6$ SPH particles. The Hill radius of the planet is resolved adequately: the radius of the planet sink is set to $0.1R_{\rm H}$ (where $R_{\rm H}$ is the Hill radius of the planet), while the number of SPH particles that is used ensures that the smoothing length, which defines the spatial resolution in SPH simulations, at the planet's initial position  is $0.17R_{\rm H}$ and $0.13R_{\rm H}$ for the simulations with $10^6$ and $2\times10^6$ particles, respectively. 
We find that the planet migrates inwards as expected with a migration timescale of $(1.7-2.6)\times 10^4$~yr when radiative transfer is included (compared to $1.5\times 10^4$ ~yr in the literature; \cite{Ayliffe:2009a}), and  $(4-7)\times 10^4$~yr for the local isothermal case, i.e. where the disc temperature depends only on the distance from the central star (compared to $(9-11)\times 10^4$~yr and  $(6-12)\times 10^4$~yr in the literature;  \cite{Ayliffe:2009a, Bate:2003a}). Therefore our calculations are in very good agreement with previous estimates of migration timescales. 

\begin{figure}[t]
\centerline{\includegraphics[width=10cm]{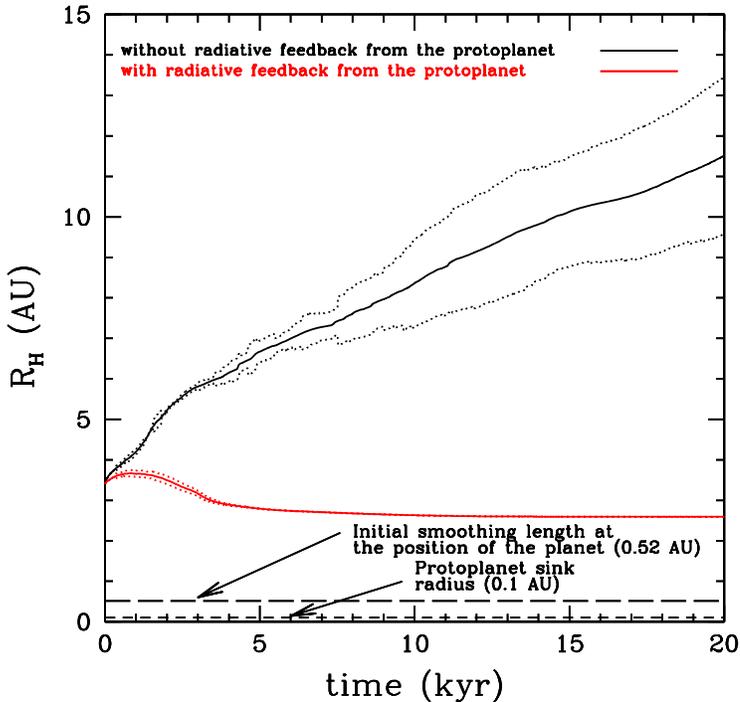}}
\caption{The evolution of the Hill radius, $R_{\rm H}$, of the protoplanet with time, compared with the smoothing length (0.52~AU), which defines the spatial resolution in SPH simulations, at the planet's initial position, and the protoplanet sink radius (0.1~AU). The Hill sphere of the protoplanet is resolved adequately.}
\label{fig:nathill}
\end{figure}

In the case of the massive, wide-orbit planets that we examine here, the planet's Hill radius is much larger and therefore easier to resolve. For example, when the Jovian protoplanet is at 50~AU away from the central star its Hill radius is $ R_{\rm H}\sim3.5$~AU. We use $10^6$ SPH particles so that the smoothing length at the initial position of the protoplanet is 0.1~AU, i.e. just $0.03R_{\rm H}$. This value remains much smaller than the Hill radius as this changes while the planet accretes mass and migrates within the disc (Figure~\ref{fig:nathill}). Another validation for our computational method  comes from the fact that the migration timescale that we obtain for the initial stage of planet migration (before a gap is opened up) is   $10^4$~yr, which in excellent agreement with previous studies \citep{Baruteau:2011a,Michael:2011a} using completely different types of codes: {\sc fargo} \citep{Baruteau:2011a} and  {\sc chymera} \citep{Michael:2011a}. 

\section{Discussion}

This work demonstrates the combined effect of gas accretion onto a protoplanet and the associated accretion-powered radiative feedback from the protoplanet, on regulating its migration and mass growth. Previous studies have shown that radiative heating from an Earth-like protoplanet forming in a low-mass disc may in fact stop its inward migration or even reverse it \citep{Benitez-Llambay:2015a}. It has also been suggested that radiative feedback may delay the contraction of the protoplanetary core so that it is tidally destroyed as it moves closer to the central star \citep{Nayakshin:2013a}.  The amount of energy radiated from the protoplanet depends on where the accretion happens, i.e. the radius of the protoplanet, which it is uncertain. It has been suggested that  this energy that is released heats the region around the protoplanet, providing a way to infer  the presence of a protoplanet in a disc \citep{Montesinos:2015a}.

Here we show that accretion and radiative feedback work in opposing ways: gas accretion increases the mass of the protoplanet and supports the creation of a gap in the disc, whereas radiative feedback, which is the result of gas accretion,  works to  suppress accretion onto the protoplanet and inhibit/delay gap opening by increasing the disc temperature. It heats and stabilises the disc, suppressing fast gas accretion. Higher gas accretion induces higher radiative feedback that in turn opposes further gas accretion. Therefore, accretion and feedback are self-regulated and the final outcome of their combined effect  is an inward migration of the protoplanet with a more moderate increase of its mass than in the case without radiative feedback. Eventually the protoplanet is able to open up a gap, the migration is effectively halted, and the protoplanet survives on a wide orbit. 

The final mass of the protoplanet in the simulation  with radiative feedback is around the deuterium burning limit. This simulation covers only a period of 20~kyr, thus if gas accretion continues, the mass of the protoplanet could further increase. However, radiative feeback may delay gas accretion until other disc dispersal mechanisms (e.g. photoevaporation) come into play.

The ramifications of these processes are important for the formation of giant planets on wide orbits. Only  a few  planets  of this type have been observed so far, but this is likely due to observational biases as these are faint objects next to bright stars and both high sensitivity and high angular resolution observations are essential. However, more wide-orbit giant planets are bound to be discovered with new surveys using specialized observing techniques  (Gemini Planet Imager \citep{Macintosh:2014b}, SPHERE/VLT \citep{Beuzit:2008a}, HiCIAO/SUBARU \citep{Suzuki:2009a}). Therefore, it is critical to explain their formation mechanism and determine their connection with the formation of  planetary systems and  the formation of our own solar system.  

Here we have showed that if massive protoplanets form by gravitational  fragmentation of discs, gas accretion and radiative feedback regulate their mass growth and inward migration,  so that they are able to avoid excessive mass growth and rapid inward migration; their mass remains around the deuterium burning limit for at least the initial stages of their evolution, they avoid disruption and  they survive on almost circular wide orbits around Sun-like stars. Therefore, it giant planets like the ones in the HR8799 system may have formed by disc fragmentation. Planet formation by  gravitational fragmentation of protostellar discs may only happen during the early stages in a disc's lifetime, on a timescale of a few thousand years, while discs are still massive enough to be prone to gravitational instabilities. The presence of a massive giant planet within the disc plays an important role in the disc's evolution and affects the dynamics of dust particles as they coagulate to form larger bodies and eventually planets within a few million years. Therefore, the formation of massive gas giant planets at an early stage and their survival on wide orbits  sets the disc environment in which subsequent formation of terrestrial planets, like Earth, happens.

\acknowledgments

I thank S. Inutsuka, H. Kobayashi, S. Ida, Y. Fujii, and M. Kunitomo for stimulating discussions on the work presented here. I also thank the anonymous referee for his/her useful comments. Simulations were performed using the UCLAN HPC Cluster {\sc wildcat}. Column density plots were produced using {\sc splash} \citep{Price:2007b}. I acknowledge support from STFC Grant ST/M000877/1 and from a Royal Society-Daiwa Foundation International Exchanges award.


\end{document}